\documentclass[a4paper,12pt]{article}
\usepackage[document]{ragged2e}
\usepackage[utf8]{inputenc}
\usepackage[T1]{fontenc} %pacote para acentuação e hifenização
\usepackage{multicol}
\usepackage{graphicx}
\usepackage{amsmath,bm}
\usepackage{textgreek}
\usepackage{booktabs}
\usepackage{MnSymbol}
\usepackage{indentfirst}
\usepackage{graphics}
\usepackage{array}
\usepackage[dvipsnames]{xcolor} % using xcolor without table option
\usepackage{float}
\usepackage{pbox}
\usepackage{wasysym}
\usepackage{threeparttable}
\usepackage{pdflscape}
\usepackage[left=2cm,right=2cm,top=25mm,columnsep=20pt]{geometry} %
\usepackage[bf,labelsep=space]{caption}
\usepackage{setspace}
\usepackage{setspace,caption}
\AtBeginCaption{\doublespacing}
\usepackage[version=3]{mhchem}
\usepackage[round,authoryear]{natbib}
\usepackage{lineno}
\usepackage{color,soul}
\usepackage{url}
\usepackage{multirow}
\usepackage{longtable}
\usepackage{lscape}
\usepackage{hyperref}
\hypersetup{
colorlinks=true, %set true if you want colored links
linktoc=all,     %set to all if you want sections and subsections linked
linkcolor=blue,  %choose some color if you want links to stand out
urlcolor=magenta, %internet links
citecolor=brown
}

\usepackage{abstract}

\usepackage{pbox}
\usepackage{booktabs}
\usepackage{mindflow}
\usepackage{authblk}
\usepackage{xspace}
\usepackage{makecell}

\usepackage[none]{hyphenat}
\makeatletter
\renewcommand\@biblabel[1]{#1.}
\makeatother
\usepackage[figurename=Fig,labelsep=period]{caption}
\usepackage{fancyhdr}
\newcommand{\covid}{COVID-19\xspace}

\begin{document}
\title{On the misuse of time-dependent models in assessing mask usage and excess mortality}
\author[$^1$]{Beny Spira}
\affil[$^1$]{Departamento de Microbiologia, Instituto de Ciências Biomédicas Universidade de São Paulo, São Paulo-SP, Brazil}
\author[$^2\ast$]{Daniel V. Tausk}
\affil[$^2$]{Departamento de Matemática, Universidade de São Paulo, São Paulo-SP, Brazil}

\affil [$ ^* $]{\small Corresponding author\\ email: tausk@ime.usp.br}
\date{}
\fancypagestyle{plain}{%
\fancyhf{}%
}

\maketitle
\pagestyle{fancy}
\renewcommand{\headrulewidth}{0pt}

\begin{abstract}
%In \citep{Tausk2025} we have shown that average mask usage during the years 2020 and 2021 is positively associated with excess mortality in that same period in $24$ European countries. Such association remains after several attempts at controlling for confounding variables. In \citep{Cerqueira-Silva2026} the authors criticize the methods used in \citep{Tausk2025} and claim that such association is explained by the fact that increased mask usage is a reaction to \covid mortality. In this article, we show that the
%analyses presented in \citep{Cerqueira-Silva2026} is fundamentally flawed and moreover that those authors mischaracterize the evidence available in the literature regarding effectiveness of masks to prevent viral transmission. Finally, we present an original analysis of the data showing that the reverse causality explanation proposed in \citep{Cerqueira-Silva2026} for the association found in \citep{Tausk2025} is probably incorrect.

The effectiveness of face masks as a population level intervention against respiratory viral transmission remains contested. While a large observational literature published during the COVID-19 pandemic reported beneficial effects, randomized controlled trials have consistently shown limited or no impact. An ecological analysis of European countries reported that average mask usage during the years 2020 and 2021 is positively associated with excess mortality in that same period in $24$ European countries \citep{Tausk2025}. Such association remains after several attempts at controlling for confounding variables.
This finding was later challenged by other authors and attributed to reverse causality \citep{Cerqueira-Silva2026}. In this paper, we reassess those criticisms in detail.
We show that their analysis is fundamentally flawed, as the time-dependent regression framework used to refute the original findings yields spurious results partly due to the use of \texttt{cumulative excess mortality} as an outcome variable, thereby incorporating pre-intervention deaths and producing statistically significant effects even at impossible negative time lags. Diagnostic analyses further demonstrate that key assumptions of the model are violated, invalidating any association or causal interpretation.
Finally, we present an original longitudinal analysis of mask usage designed to directly test the reverse causality hypothesis. By constructing multiple indices that capture mask adoption during distinct phases of pandemic waves, including interwave periods characterized by low mortality, we show that the association between mask usage and excess mortality persists and is not driven by reactive increases in masking. These findings provide substantial evidence that reverse causality provides, at most, a minor contribution to the observed association. The results highlight the importance of appropriate outcome definitions, model validation, and evidentiary rigour in the evaluation of non-pharmaceutical interventions.
\end{abstract}

\section{Introduction}

The use of face masks as a tool to limit the spread of respiratory viruses remains controversial. On the one hand, the  majority of papers published during the \covid\ pandemic adopted a  pro-mask stance and claimed that masks are effective in preventing viral transmission \citep{Spira2026}. Mask usage was also recommended by the World Health Organization (WHO) and mandated in many countries during this period \citep{Masks4All2020}. In addition, many individuals, including health care workers and scientists, intuitively believed that masks offered personal protection against \covid\ infection.

On the other hand, the highest-quality available evidence -- randomised controlled trials (RCTs), has consistently shown that masks have little to no effect on viral transmission. These studies are summarised in \autoref{RCTs}. Among the 16 RCTs that have evaluated mask usage \citep{Cowling2008,Cowling2009,Canini2010,Larson2010,Simmerman2011,Suess2012,MacIntyre2016,Aiello2012,Barasheed2014,Alfelali2020,Bundgaard2020,Solberg2024,Abaluck2022,Nanque2024,Jacobs2009,MacIntyre2015}, only two reported statistically significant positive results \citep{Solberg2024,Abaluck2022}, and even these reported modest effects and were accompanied by substantial limitations.

\begin{table}[H]
\small
\caption{Randomized controlled trials of face masks to date}
\label{RCTs}
\begin{tabular}{|m{3cm}|m{2.2cm}|m{1cm}|m{1.7cm}|m{3cm}|m{5cm}|}
\hline
\textbf{Author, year, country}  & \textbf{Setting} & \textbf{N}  & \textbf{Primary endpoint} & \textbf{Risk ratio (95\% CI)} & \textbf{Comment}  \\
\hline
Cowling et al. (2008) Hong Kong  & Community & 198 & ILI & 0.88 [0.34 , 2.27] & medical/surgical masks versus no masks \\
 &  &  & LCI & 1.16 [0.31 , 4.34] &  \\
\hline
Canini et al. (2010) France  & Community & 105 & ILI & 1.03 [0.52 , 2.00] & medical/surgical masks versus no masks \\
\hline
Larson et al. (2010) USA  & Community & 509 & ILI & 0.63 [0.31 , 1.29] & hand hygiene + medical/ surgical masks compared to hand hygiene \\
 &  &  & LCI & 0.57 [0.17 , 1.87] &  \\
\hline
Simmerman et al. (2011) Thailand  & Community & 442 & ILI & 1.03 [0.61 , 1.73] & hand hygiene + medical/ surgical masks compared to hand hygiene \\
 &  &  & LCI & 0.97 [0.62 , 1.52] &  \\
\hline
Suess et al. (2012) Germany  & Community & 84 & ILI & 0.61 [0.20 , 1.87] & medical/surgical masks versus no masks \\
 &  &  & LCI & 0.39 [0.13 , 1.19] &  \\
\hline
MacIntyre. (2016) China et  & Community & 245 & ILI & 0.32 [0.03 , 3.11] & medical/surgical masks versus no masks \\
\hline
Aiello et al. (2010) USA  & Community & 1437 & LCI & 2.43 [0.59, 10.12] & medical/surgical masks versus no masks \\
 &  &  & ILI & 0.82 [0.65, 1.01] &  \\
\hline
Aiello et al. (2012) USA  & Community & 1178 & ILI & 1.10 [0.88 , 1.38] & medical/surgical masks versus no masks \\
 &  &  & LCI & 0.92 [0.59 , 1.42] &  \\
\hline
Barasheed et al. (2014) Saudi Arabia  & Community & 164 & ILI & 0.58 [0.32 , 1.04] & medical/surgical masks versus no masks \\
\hline
Alfelali et al. (2020) Saudi Arabia  & Community & 7687 & ILI & 1.10 [0.90 , 1.35] & medical/surgical masks versus no masks \\
 &  &  & LCI & 1.40 [0.92 , 2.14] &  \\
\hline
Bundgaard et al. (2021) Denmark   & Community & 4862 & LCI* & 0.82 [0.54 , 1.23] & medical/surgical masks versus no masks \\

 &  &  & LCI** & 0.58 [0.25 , 1.31] &  \\
\hline
Solberg et al (2024) Norway  & Community & 4467 & SRRS & 0.71 [0.58-0.87] & medical/surgical masks versus no masks. \\
\hline
Abaluck et al. (2022) Bangladesh  & Community & 336010 & CLI & 0.87 [0.81 , 0.94] & medical/surgical masks versus no masks \\
\hline
Nanque et al. (2024) Guinea- Bissau  & Community & 39574 & CLI & 0.81 [0.57, 1.15] & cloth masks versus no masks \\
\hline
Jacobs et al. (2009) Japan  & Hospital & 33 & ILI &  0.88, [0.02, 32] & medical/surgical masks versus no masks \\
\hline
MacIntyre et al. (2015) Vietnam  & Hospital & 1607 & ILI & 0.26 [0.03, 2.50] & medical masks versus cloth masks \\
\hline
\end{tabular}
CLI, COVID-like illness; ILI, Influenza-like illness; LCI, Lab-confirmed infection; LCI*, Lab-confirmed SARS-CoV-2 infection; LCI**, Lab-confirmed (other viruses); SRRS, Self-reported respiratory symptoms.
\end{table}

The community-based RCT conducted by \citet{Solberg2024} reported a 29\% relative reduction in the risk of respiratory infection associated with surgical mask usage in public spaces; however, this corresponded to only a 3.2\% absolute risk reduction. Moreover, the primary endpoint was broad and subjective, namely `self-reported respiratory symptoms'. When the outcome was restricted to the more rigorous \covid\ confirmation by PCR or antigen testing, the estimated odds ratio was 1.07 and not statistically significant. Additionally, subgroup analyses suggested an apparent benefit only among participants who reported believing that masks reduce infection.

The large cluster-randomised trial by \citet{Abaluck2022} reported a statistically significant reduction in COVID-19-like illness among mask users (a relative risk reduction of 13\%)%RR 0.87, 95\% CI 0.81 to 0.94)
. However, this study was rated as being at high risk of bias in five of six domains in the subsequent Cochrane review on physical interventions to reduce the spread of respiratory viruses \citep{Jefferson2023}, due to issues including baseline imbalances, subjective outcome assessment, incomplete follow-up, and deviations from the pre-specified protocol. Notably, baseline serological testing specified in the protocol was not performed, and no adequate justification was provided.

%Finally, the RCT by \citet{MacIntyre2015}, conducted in a healthcare setting, found that medical or surgical masks were more effective than cloth masks in reducing the rate of influenza-like illness (RR 13.25, 95\% CI 1.74 to 100.97). However, the extremely wide confidence intervals render this estimate highly imprecise and difficult to interpret, limiting its evidentiary value.
Taken together, the higher-quality randomised studies provide little support for a substantial protective effect of masks against respiratory viral transmission (see Subsection~\ref{sub:Cochrane}), standing in marked contrast to the overwhelmingly positive portrayal of masks in the broader pandemic literature.

The RCTs evidence did not lessen the torrent of articles portraying masks as relevant and effective tools in the fight against viral infections \citep{Spira2026}. Nevertheless, a small number of papers, that were neither RCTs nor their reviews or meta-analyses, have challenged this hegemonic narrative. One such study was an ecological observational analysis of the relationship between mask usage and excess mortality in Europe \citep{Tausk2025}. There, bivariate and multivariate regression analyses were used to examine correlations between average mask usage and various excess mortality measures across 24 European countries. Statistically significant positive correlations were identified between mask usage rates and age-adjusted excess mortality in both bivariate analyses (Spearman’s $\rho = 0.477$, $p = 0.018$) and multivariate models (standardised coefficient $= 0.52$, $p = 0.0012$). In contrast, vaccination rates were negatively and significantly associated with age-adjusted excess mortality in both bivariate (Spearman’s $\rho = -0.659$, $p < 0.001$) and multivariate analyses (standardised coefficient $= -0.48$, $p = 0.0016$), consistent with a protective effect of vaccination. Extensive consideration was given to potential confounders, including socioeconomic factors, indicators of baseline population health, urban population density, vaccination rates, and the stringency of other non-pharmaceutical interventions. In addition, multiple sensitivity analyses were conducted to assess the robustness of the findings. The statistically significant positive association between mask usage and excess mortality persisted in all analyses considered.

Following publication, the journal \textit{BMC Public Health} was subjected to pressure to retract the article \citep{jusp2026}. This led Springer Nature’s research integrity office to initiate a post-publication review, which concluded several months later with the publication of an \textit{erratum} \citep{Tausk2025a}. The \textit{erratum} consisted of some methodological clarifications and minor revisions to three sentences in the Discussion section, adopting more cautious language without altering the study’s results or conclusions.

Subsequently, and apparently unsatisfied with this outcome, a group of authors published an article in \textit{The Lancet Regional Health -- Americas} \citep{Cerqueira-Silva2026} with the explicit aim of undermining the findings of \citet{Tausk2025}.

The authors argued that the purported link between mask usage and excess mortality is based on flawed methodological assumptions, particularly the use of aggregate averages that ignore temporal fluctuations in mask adoption. They contend that mask usage is actually a reactive response to worsening outbreaks and increased government stringency. To counter \cite{Tausk2025}, the authors performed a time-dependent analysis of the same data using a two-way fixed effects model which revealed a %positive non statistically significant association of mask usage and \texttt{weekly excess mortality} and a
statistically significant negative association between mask usage and \texttt{cumulative excess mortality}.
The authors also argued that the original findings lack clinical and biological plausibility because, according to them, masks are scientifically proven barriers to viral transmission.
Furthermore, their critique is framed through a political lens, as they invoke the study’s alleged use in denialist rhetoric, thereby politicising the discussion rather than keeping it strictly scientific.

In the present paper, we rebut the claims of \citet{Cerqueira-Silva2026} by showing that the two-way fixed effects model they used to assess the relationship between mask usage and excess mortality is fundamentally flawed. We also address additional important mischaracterisations present in their article. Finally, we present an original analysis showing that the positive association between mask usage and excess mortality reported in \citet{Tausk2025} is unlikely to be explained by reverse causality, that is, by increases in mask usage occurring in response to higher \covid mortality during the pandemic.

\section{Results and Discussion}

\subsection{The two-way fixed effects model used by \citet{Cerqueira-Silva2026} is demonstrably flawed}\label{sub:twowayflawed}

\citet{Cerqueira-Silva2026} used a two-way fixed effects model to estimate the effect of mask usage on the excess mortality and on the \texttt{cumulative excess mortality} in the subsequent weeks. More precisely, for {every week} $t$ between January 1st 2020 and January 2nd 2022, the coefficient $\beta$ of their model measures the effect of the average mask usage on week $t$ on the excess mortality on week $t+d$ and on the \texttt{cumulative excess mortality} between January 1st 2020 and week $t+d$, where the lag $d$ ranges from 1 to 4 {(see \citet{owid_data} for the definition of the \texttt{cumulative excess mortality} variable)}. Point estimates of $\beta$ and corresponding confidence intervals are reported in their Supplement section, which is reproduced below for convenience (\autoref{Reproduced from Silva et al.}).

\begin{table}[H]
\centering
\caption{Reproduced from \citet{Cerqueira-Silva2026}}
\label{Reproduced from Silva et al.}
\begin{tabular}{lcc}
\hline
Lag & Weekly excess mortality $\beta$ (95\% CI) & Cumulative excess mortality $\beta$ (95\% CI) \\
\hline
1 week  & 0.0524 [$-0.1008$; 0.2057] & $-0.0674$ [$-0.1270$; $-0.0077$] \\
2 weeks & 0.0426 [$-0.1015$; 0.1867] & $-0.0688$ [$-0.1269$; $-0.0107$] \\
3 weeks & 0.0269 [$-0.1107$; 0.1644] & $-0.0702$ [$-0.1263$; $-0.0140$] \\
4 weeks & 0.0134 [$-0.1207$; 0.1476] & $-0.0713$ [$-0.1253$; $-0.0172$] \\
\hline
\end{tabular}
\end{table}

If the goal is to study the effect of mask usage on mortality in a time-dependent manner, the natural choice of outcome variable is \texttt{weekly excess mortality} measured a certain number of weeks after the period of mask usage, i.e., excess mortality during the time window in which mask usage could plausibly exert an effect. The \texttt{cumulative excess mortality} is a somewhat odd choice of outcome variable as it encompasses the excess mortality of a large (and varying) number of weeks before the period of mask usage. It seems that the authors understand that as, in their own words, \texttt{cumulative excess mortality} ``{\em may introduce distinct biases by smoothing over heterogeneous epidemic phases}''.

We note that when \texttt{weekly excess mortality} is used as outcome variable, the point estimates of the parameter $\beta$ are positive for all lags. This is consistent with our findings that more mask usage is associated with more excess mortality. However, since the result is not statistically significant, one cannot infer much from this.

It is only when the \texttt{cumulative excess mortality} is used as an outcome variable that estimates of $\beta$ become negative and statistically significant, indicating that mask usage is reducing excess mortality. However, the observation of an effect with a lag of only one week is a red flag, since if masks reduce transmission, any impact on mortality would be expected to manifest only after a longer time span. Since estimates of the interval between cases and deaths vary in the literature \citep{Fritz2022}, we decided to settle the matter by using the code supplied by \citet{Cerqueira-Silva2026} to compute the estimates of the parameter $\beta$ for lags equal to 0 and $-1$ (\autoref{Extended lag analysis}).

\begin{table}[H]
\centering
\caption{Extended lag analysis}
\label{Extended lag analysis}
\begin{tabular}{lcc}
\hline
Lag & Weekly excess mortality $\beta$ (95\% CI) & Cumulative excess mortality $\beta$ (95\% CI) \\
\hline
$\bm{-1}$ \textbf{week} & 0.0438 [$-0.1246; 0.2123$] & $\bm{-0.0639}$ [$\bm{-0.1251}; -\bm{0.0027}$] \\
\textbf{0 weeks}  & 0.0523 [$-0.1093; 0.2140$] & $\bm{-0.0658}$ [$\bm{-0.1264}; \bm{-0.0051}$] \\
1 week   & 0.0524 [$-0.1008; 0.2057$] & $-0.0674$ [$-0.1270; -0.0077$] \\
2 weeks  & 0.0426 [$-0.1015; 0.1867$] & $-0.0688$ [$-0.1269; -0.0107$] \\
3 weeks  & 0.0269 [$-0.1107; 0.1644$] & $-0.0702$ [$-0.1263; -0.0140$] \\
4 weeks  & 0.0134 [$-0.1207; 0.1476$] & $-0.0713$ [$-0.1253; -0.0172$] \\
\hline
\end{tabular}
\end{table}

\autoref{Extended lag analysis} shows that the estimates of $\beta$ and corresponding confidence intervals obtained for lags of $0$ and $-1$ are very similar to the estimates obtained for lags between 1 and 4 weeks. Since it is implausible that mask usage in a given week could affect \texttt{cumulative excess mortality} in that same week, and impossible that it could influence excess mortality in preceding weeks, this result conclusively demonstrates that the effects reported by \citet{Cerqueira-Silva2026} are artifacts arising from the inclusion of pre-intervention mortality in the outcome variable. The reported effects are therefore spurious and render the analysis fundamentally unsound.
In addition, we show below that important assumptions of the model used by \citet{Cerqueira-Silva2026} are violated. This might explain why the model predicts a spurious effect of mask usage on excess mortality during the pre-masking period.

\subsubsection{The assumptions underlying the model are invalid}

The two-way fixed effects model used by \citet{Cerqueira-Silva2026} has the following form
\[
Y_{it+d} = a_i + b_t + \beta M_{it} + \varepsilon_{it},
\]
where $Y_{it}$ is the outcome variable (either excess mortality or \texttt{cumulative excess mortality}) for country $i$ on week $t$, $d$ is the fixed chosen time lag between 1 and 4 weeks, $a_i$ is a parameter that represents country-specific time-independent effects, $b_t$ is a parameter that represents time-specific country-independent effects, $\beta$ is the parameter of interest, $M_{it}$ is the average mask usage for country $i$ on week $t$ and $\varepsilon_{it}$ is a random error term. In short, aside from the effect of mask usage and random noise, the outcome variable is the sum of a country-specific time-independent term $a_i$ with a time-specific country-independent term $b_t$.

It is well known that the validity of the additive structure $a_i + b_t$ of the main effect is crucial for an adequate estimation of the parameters of interest in a two-way fixed effects model \citep{Imai2020}. In fact, it is easy to prove that if this additivity assumption fails then the usual ordinary least squares (OLS) estimator for $\beta$ is often biased and the bias does not tend to zero as sample size goes to infinity, i.e., the estimator is inconsistent.

In the case at hand, it is clear that the additivity assumption is unrealistic. At the very least, time lags between infection waves in different countries make the time effects country dependent. Moreover, country-dependent effects are not in general expected to act additively. Small differences in viral reproductive numbers, for instance, would generate complicated non-linear effects on infection waves. Also, differences in age structure, general baseline health of the population and quality of health infrastructure would change the {infection fatality rate} generating multiplicative country-specific effects.

In order to assess the validity of the model assumptions, we plotted residuals against time for each of the 24 countries. This is a standard graphical diagnostic for linear models: when the assumptions are satisfied, residuals should behave like random noise fluctuating around zero. In the present case, the opposite is observed. The 24 residual plots using \texttt{cumulative excess mortality} as the outcome variable and time lags $d$ between $1$ and $4$ weeks are shown in the files available in a GitHub repository \citep{Tausk2026}. These plots provide conclusive evidence that the model assumptions are severely violated. Had \citet{Cerqueira-Silva2026} performed this basic diagnostic check, they would have seen that their model is fundamentally flawed.

\subsection{Addressing the reverse causality problem}\label{sub:reverse}

One of the central claims advanced by \citet{Cerqueira-Silva2026} is that countries exhibiting higher levels of mask usage did so in response to more severe pandemic waves and that this fact explains the positive association between mask usage and excess mortality found in \citet{Tausk2025}. We call this objection the {\em reverse causality problem}: the causal direction (if any) is not from the intervention (mask usage) to the outcome (mortality), but the opposite. Figure~1d in \citet{Cerqueira-Silva2026} is presented as the sole piece of direct evidence supporting the assertion that mask adoption increased reactively following worsening outbreaks. As they state:

\begin{quotation}
	\textit{To further contextualise the findings, we included data from Brazil (Fig.~1d) showing the excess mortality (monthly) and mask usage (30-day moving average). The figure suggests that an increase in mask adoption typically occurs after peaks in mortality, indicating a reactive response to worsening outbreaks rather than a causal effect, contradicting Tausk and Spira’s conclusions.}
\end{quotation}

However, in addition to the fact that the chart represents data from a single country -- Brazil, and not from Europe as would be expected from a paper criticising a study made on European countries, it is difficult to see how this chart supports the claim of reverse causality. For instance, mask usage rate during the second wave, peaking in April 2021, was virtually identical to that during the first wave in June 2020, remaining near 70\%, despite the second wave being far more lethal, with excess mortality of approximately 80\% compared with about 27\% in the first wave. If reverse causality were operating, one would expect substantially higher mask usage during the second wave, yet no such increase was observed. The data presented by \citet{Cerqueira-Silva2026} therefore refute, rather than support, their claim of reverse causality, undermining the credibility of their interpretation.

Though the arguments presented in \citet{Cerqueira-Silva2026} for reverse causality are invalid, we recognize that the reverse causality problem is a reasonable concern and that it was not sufficiently addressed in our original article. We therefore address this issue in detail here.

We start by defining five indices -- \texttt{maskall}, \texttt{maskinwave}, \texttt{maskinterwave}, \texttt{maskbeginwave} and \texttt{maskpeakwave\/} which correspond to measures of mask usage in different time periods. We note that, since the pandemic waves do not start and end at the same time in different countries, in a time-dependent analysis it makes little sense to compare masking and mortality across countries at the same time periods. For this reason, the time periods associated to the five indices are defined in terms of their relationship with the pandemic waves and not in terms of calendar time. Three pandemic waves were identified for each country (\autoref{gph:phases}). The precise methodology used to define the onset and termination of each pandemic wave within a country is described in the Methods section.
Below we define the five mask-usage indices:

\begin{itemize}
\item \texttt{maskall}: average mask usage during the entire period of analysis (from 2020/02/04 to 2021/12/31);
\item \texttt{maskinwave}: average mask usage during the periods corresponding to pandemic waves;
\item \texttt{maskinterwave}: average mask usage during interwave periods, defined as the complement of the periods corresponding to \texttt{maskinwave};
\item \texttt{maskbeginwave}: average mask usage during the early phase of pandemic waves, defined as the first decile of the wave's duration;
\item \texttt{maskpeakwave}: average mask usage around the peak of pandemic waves, defined as the decile of a wave starting on the day with the maximum COVID-19 mortality during that wave.
\end{itemize}

The index \texttt{maskall} is essentially identical to that used in \citet{Tausk2025} (the only difference is that in that study zero values at the beginning of the pandemic were excluded before computing the mean). We assume that any influence of mortality on mask usage is local in time, meaning that masking in a given time period can only be influenced by mortality in the recent past.
Under this assumption, the index \texttt{maskinterwave} can only be influenced by mortality occurring near interwave periods which is very low compared with mortality during mid-wave periods. Consequently, the portion of mortality that could plausibly affect \texttt{maskinterwave} carries little weight in \covid mortality over the entire period of analysis, implying that reverse causality can play at most a minor role in generating correlations between \texttt{maskinterwave} and \covid mortality over the full period.
By contrast, reverse causality could plausibly play a substantial role in generating correlations between \covid mortality and the indices \texttt{maskall}, \texttt{maskinwave} and \texttt{maskpeakwave}. The index \texttt{maskbeginwave} occupies an intermediate position between \texttt{maskinterwave} and these indices: reverse causality could contribute more to correlations involving \texttt{maskbeginwave} than to those involving \texttt{maskinterwave}, but less than to those involving the other indices.

Once the relevant mask usage indices were defined, we examined their interrelationships by computing pairwise Pearson correlation coefficients.

\begin{minipage}{\textwidth}
\small
\begin{longtable}{|cc|}
%\centering
\caption{Pearson correlations and corresponding 95\% confidence intervals between mask usage indices.}
\label{tab:pearsonmaskindices}\\

\toprule
\textbf{Indices} & \textbf{Correlation}  \\

\midrule[0.4pt]

\texttt{maskall} and \texttt{maskinterwave}&$0.95\ [0.92,1]$\\
\texttt{maskall} and \texttt{maskinwave}&$0.99\ [0.99,1]$\\
\texttt{maskall} and \texttt{maskbeginwave}&$0.94\ [0.90,1]$\\
\texttt{maskall} and \texttt{maskpeakwave}&$0.94\ [0.91,1]$\\
\texttt{maskinterwave} and \texttt{maskinwave}&$0.93\ [0.89,1]$\\
\texttt{maskinterwave} and \texttt{maskbeginwave}&$0.92\ [0.87,1]$\\
\texttt{maskinterwave} and \texttt{maskpeakwave}&$0.86\ [0.79,0.98]$\\
\texttt{maskinwave} and \texttt{maskbeginwave}&$0.96\ [0.93,0.99]$\\
\texttt{maskinwave} and \texttt{maskpeakwave}&$0.96\ [0.94,1]$\\
\texttt{maskbeginwave} and \texttt{maskpeakwave}&$0.92\ [0.87,0.98]$\\

\bottomrule
\end{longtable}
\normalsize
\vspace{0.5cm}
\end{minipage}

Table~\ref{tab:pearsonmaskindices} shows that the pairwise Pearson correlations are close to $1$, indicating that these indices are approximately affine increasing transformations of one another (i.e., transformations of the form $y=ax+b$ with $a>0$). In simpler terms, these various measures of the amount of mask usage rank countries in similar ways and switching from one index to the other is akin to a mere change of scale.

The high values of Pearson correlations also imply that these indices are largely interchangeable for the purpose of multiple regression analysis. In our original article, a multiple regression analysis was performed using age-adjusted excess mortality as the outcome variable and vaccination rate, mask usage percentage, human development index and the principal component of cardiovascular death rate and life expectancy as explanatory variables (See Table~3 in \citet{Tausk2025}; additional regression models were examined in sensitivity analyses and are reported in Supplementary Material~2 of that article.).
Table~\ref{tab:mainreg} reports the results of the multiple regression used in our original article, with the mask variable replaced in turn by each of the five indices defined above. For each mask index, the standardized regression coefficient of the mask variable, along with its corresponding $95\%$ confidence interval and p-value are shown. As expected, the results are similar across all mask usage indices.

\begin{minipage}{\textwidth}
\small
\begin{longtable}{|ccc|}
%\centering
\caption{Replication of the multiple regression analysis from \citet{Tausk2025}, replacing the mask variable with each of the five mask-usage indices.}
\label{tab:mainreg}\\

\toprule
\textbf{Mask index} & \textbf{standardized coefficient [CI]} & \textbf{p-value} \\

\midrule[0.4pt]

\texttt{maskall}&$0.51\ [0.23,0.79]$&$0.001$\\
\texttt{maskinterwave}&$0.44\ [0.16,0.72]$&$0.005$\\
\texttt{maskinwave}&$0.51\ [0.23,0.78]$&$0.001$\\
\texttt{maskbeginwave}&$0.51\ [0.25,0.78]$&$0.001$\\
\texttt{maskpeakwave}&$0.52\ [0.28,0.76]$&$0.0002$\\

\bottomrule
\end{longtable}
\normalsize
\vspace{0.5cm}
\end{minipage}

For the reasons discussed above, if the association between mask usage and mortality reported in \citet{Tausk2025} were driven by reverse causality, that association should largely disappear once the mask variable in the multiple regression is replaced by the \texttt{maskinterwave} index. In addition, one would expect the association involving \texttt{maskbeginwave} to be substantially weaker than the original association between excess mortality and overall mask usage. However, what we observe is that the standardized coefficient is only slightly smaller when using the \texttt{maskinterwave} index and the p-value remains very low. Moreover, replacing \texttt{maskall} with \texttt{maskbeginwave} produces virtually no change in the results. This shows that at most a small fraction of the observed association between mask usage and mortality can be attributed to reverse causality.

Next, we searched for direct evidence of reverse       causality. To this end, we performed additional comparisons among the mask usage indices and examined how the relationships between these indices varied with COVID-19 mortality. First, we compared \texttt{maskbeginwave} with \texttt{maskpeakwave}. In this analysis we consider waves separately, so that we have $3\times24=72$ datapoints (three waves for each of the $24$ countries). In this setting, Pearson correlation is $0.85$ ($95\%$ CI $[0.80,0.93]$). The index \texttt{maskpeakwave} is larger than \texttt{maskbeginwave} ($p < 0.0001$ by one-sided paired Wilcoxon test), i.e., mask usage is larger near the peak of a wave than near the beginning of a wave.

The high Pearson correlation between \texttt{maskbeginwave} and \texttt{maskpeakwave} means that the scatter plot of these two variables is well-approximated by the corresponding regression line. Neglecting, in a first approximation, the intercept of the regression line, this observation motivates expressing the effect of the pandemic wave on mask usage by the ratio of \texttt{maskpeakwave} by \texttt{maskbeginwave}, which can be interpreted as an estimate of the slope of the regression line.
The median of this ratio is $1.52$ with the first and third quartiles of $1.19$ and $2.12$, respectively.

\begin{figure}[H]
\centering 	
\includegraphics[width=\linewidth]{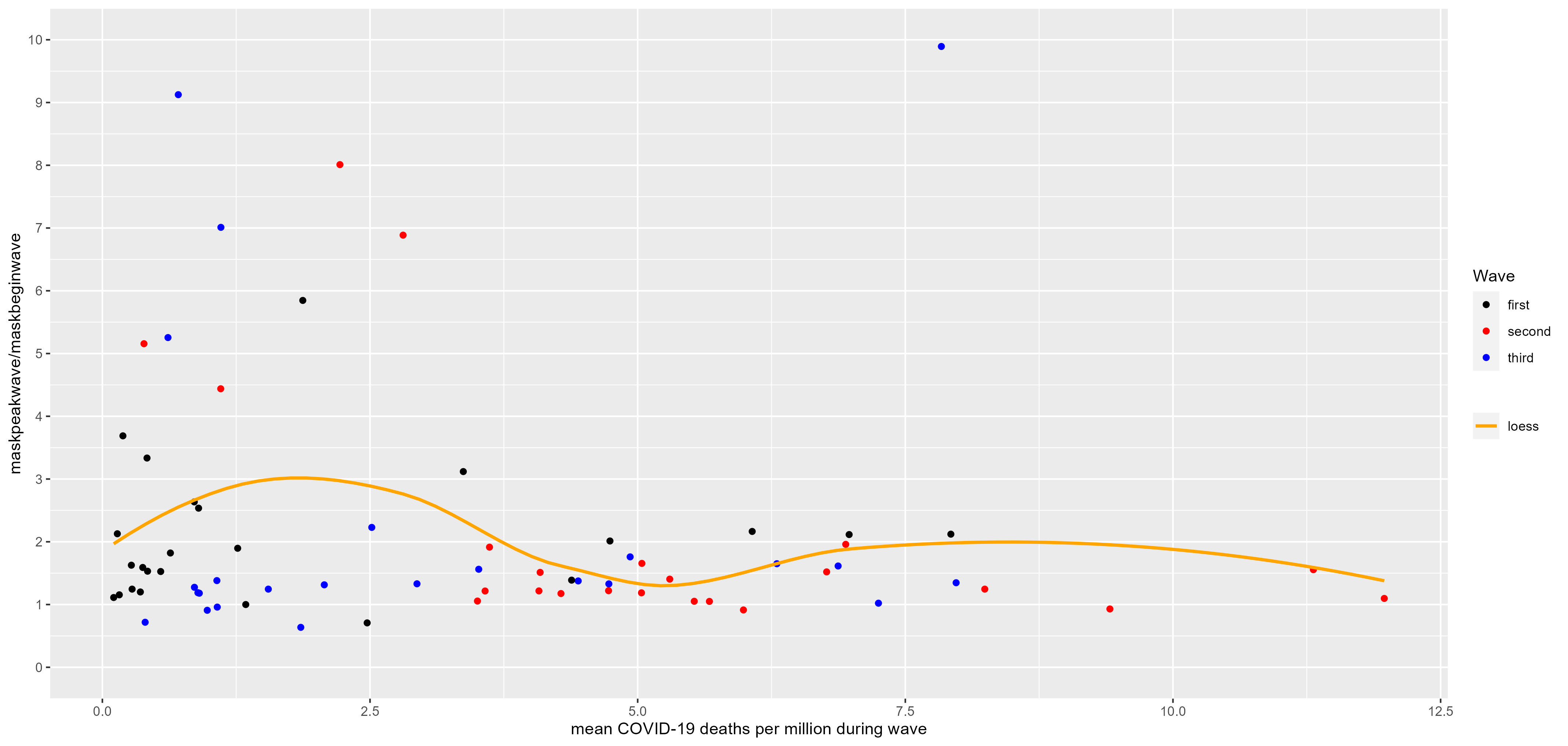}
\caption{scatter plot of the ratios \texttt{maskpeakwave}/\texttt{maskbeginwave} against mean COVID-19 mortality per million during the corresponding wave.}
\label{gph:peakbegin}
\end{figure}

\autoref{gph:peakbegin} shows a scatter plot of the ratio \texttt{maskpeakwave}/\texttt{maskbeginwave} against mean COVID-19 mortality per million during the corresponding wave. Spearman correlation is slightly negative and not statistically significant $-0.13$ ($95\%$ CI $[-0.37,0.09]$), suggesting that the increase in mask wearing during a wave is not influenced by the amount of COVID-19 mortality per million during that wave.

This issue can alternatively be examined by regressing \texttt{maskpeakwave} on both \texttt{maskbeginwave} and the mean COVID-19 mortality per million during the wave. Here, the coefficient of the mean COVID-19 mortality represents the effect of COVID-19 mortality on \texttt{maskpeakwave}, when we control for \texttt{maskbeginwave}. The standardized coefficient is $0.03$ with a $95\%$ confidence interval $[-0.10,0.15]$ ($\text{p}=0.69$). This provides evidence that, once baseline mask usage is accounted for,  increases in mask wearing during a wave are not influenced by the magnitude of COVID-19 mortality during that wave.

\begin{figure}[H]
\centering 	
\includegraphics[width=\linewidth]{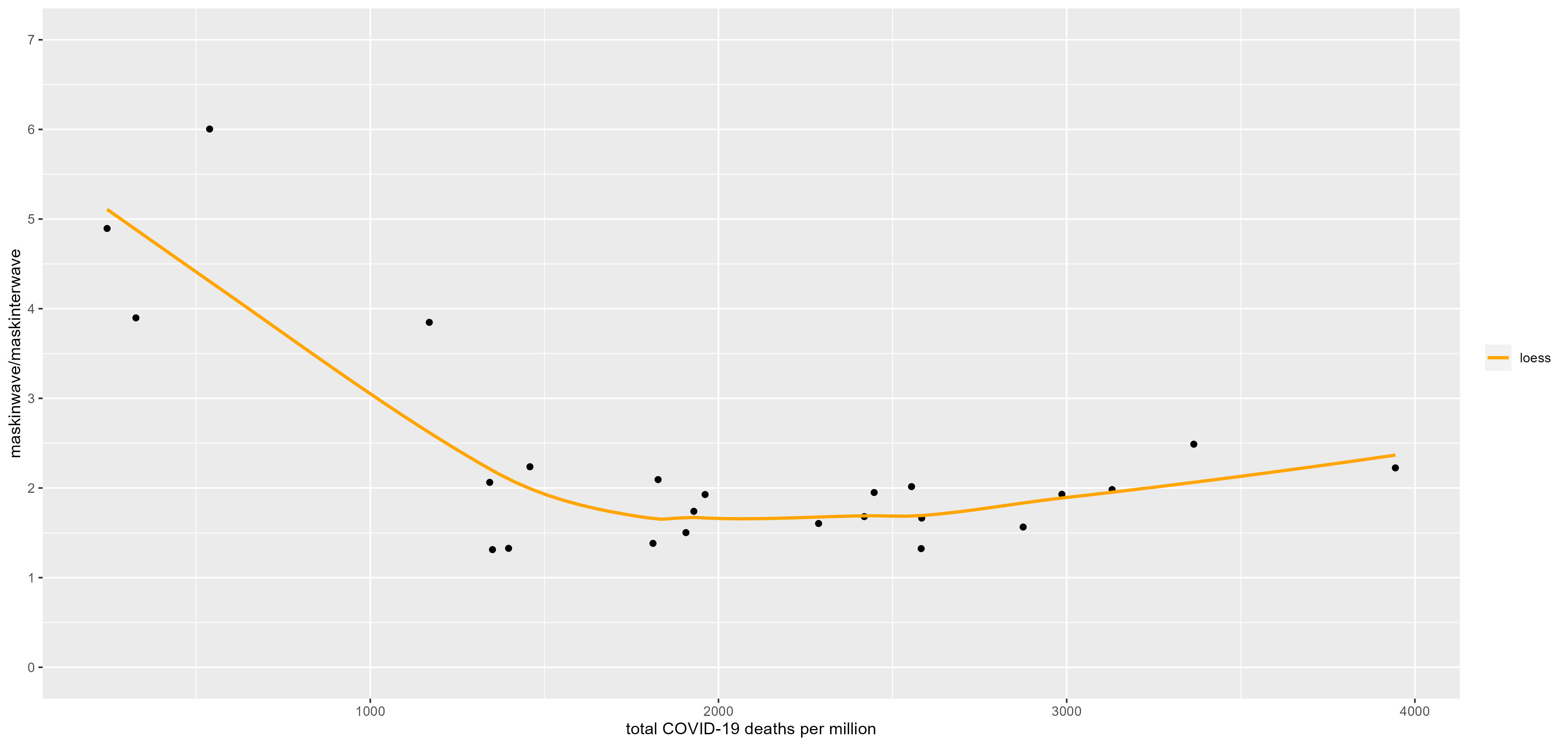}
\caption{Scatter plot of the ratio \texttt{maskinwave}/\texttt{maskinterwave} against total COVID-19 mortality per million over the entire period of analysis.}
\label{gph:ininter}
\end{figure}

A similar analysis can be performed by comparing \texttt{maskinwave} with \texttt{maskinterwave} (considering here one datapoint per country). The index \texttt{maskinwave} is larger than \texttt{maskinterwave} ($p < 0.0001$ by one-sided paired Wilcoxon test). The median of the ratio \texttt{maskinwave}/\texttt{maskinterwave} is $1.94$ and the first and third quartiles are $1.59$ and $2.23$, respectively.
\autoref{gph:ininter} shows a scatter plot of \texttt{maskinwave}/\texttt{maskinterwave} against  COVID-19 deaths per million over the entire period of analysis. Spearman correlation is negative and not statistically significant $-0.25$ ($95\%$ CI $[-0.81,0.17]$). Regressing \texttt{maskinwave} on both \texttt{maskinterwave} and total COVID-19 mortality per million results in a standardized coefficient of the mortality variable of $0.11$ with a $95\%$ confidence interval $[-0.06,0.28]$ ($\text{p}=0.20$).
Thus, there is very little evidence that increases in mask wearing during pandemic waves are influenced by the level of COVID-19 mortality.

In conclusion, while mask usage does increase during the more critical moments of the pandemic, being higher during waves than between waves and higher around wave peaks than at their onset, cross-country and cross-wave comparisons indicate that the magnitude of this increase is unlikely to be related to the level of COVID-19 mortality. Additionally, a previous study examining mask usage during the second \covid wave in the 2020-2021 winter reported a positive correlation between average mask usage and \covid mortality \citep{Spira2022}. Crucially, when this second wave began, mask mandates and their enforcement had already been in place for several months, having been introduced in the first half of 2020. These measures were strictly enforced in countries such as France, Germany, Italy, Portugal, and Spain, while enforcement was comparatively weaker in Scandinavian countries. Mask use therefore clearly preceded the subsequent rise in infections and mortality during the winter wave, rather than emerging as a reaction to it.
Altogether, these findings provide compelling evidence that reverse causality does not account for the positive association between mask usage and mortality reported in our original article.

\subsection{Misrepresentation of the Cochrane review’s key findings on masks}\label{sub:Cochrane}

\citet{Cerqueira-Silva2026} state that

\begin{quotation}
\textit{Substantial evidence shows that masks, especially high-grade masks used correctly, are effective barriers to viral transmission or, at worst, do not increase infections or mortality.\textsuperscript{6--11}}
\end{quotation}

One of the references cited in support of this claim is the Cochrane review on non-pharmaceutical interventions (ref.~no.~8). However, the Cochrane authors explicitly concluded that:

\begin{quotation}
\textit{The pooled results of RCTs did not show a clear reduction in respiratory viral infection with the use of medical-surgical masks. There were no clear differences between the use of medical-surgical masks compared with N95-P2 respirators in healthcare workers when used in routine care to reduce respiratory viral infection.}
\end{quotation}

It is therefore unclear how this review could be cited as evidence of mask effectiveness. Moreover, the conclusions regarding masks in the 2023 review are qualitatively identical to those of the previous Cochrane review published in late 2020 \citep{Jefferson2020}, which reached essentially the same assessment. It is understandable why \citet{Cerqueira-Silva2026} sought to include a Cochrane review among the studies purportedly supporting their thesis, given that Cochrane reviews are internationally recognised as the gold standard for high-quality evidence on the effectiveness of healthcare interventions.
However, by portraying the Cochrane review as supportive of mask usage for reducing viral transmission, \citet{Cerqueira-Silva2026} fundamentally mischaracterise its conclusions and thereby mislead readers about the actual state of the evidence.
Additionally, the claim that masks clearly do not increase infections, as stated in the second part of their sentence, is not entirely true. The confidence intervals for all three outcomes analysed in the Cochrane systematic review \citep{Jefferson2023} included values above 1.0. Specifically, the pooled risk ratios ([95\%,\text{CI}]) were 0.95 ([0.84,1.09]) for influenza or COVID-like illness, 1.01 ([0.72,1.42]) for laboratory-confirmed influenza or SARS-CoV-2 infection, and 0.58 ([0.25,1.31]) for laboratory-confirmed infection with other respiratory viruses. Moreover, several RCTs included in the Cochrane review reported higher infection rates in the mask groups, i.e., direction of effect is not consistent throughout studies (see \autoref{RCTs}).

\section{Methods}

Pearson correlations were used for pairwise comparisons of variables whenever it was relevant to determine if they were approximately related by an affine transformation $y=ax+b$. Otherwise, Spearman correlations were used. In all cases, confidence intervals were computed using basic bootstrap (thus avoiding bivariate normality assumptions).
For multiple regressions, confidence intervals for standardized regression coefficients were computed using the betaDelta R package (assuming multivariate normal approximation)
and p-values were computed for non standardized coefficients (since the null hypothesis is the same for standardized and non standardized coefficients).

\begin{figure}[H]
\centering 	
\includegraphics[width=\linewidth]{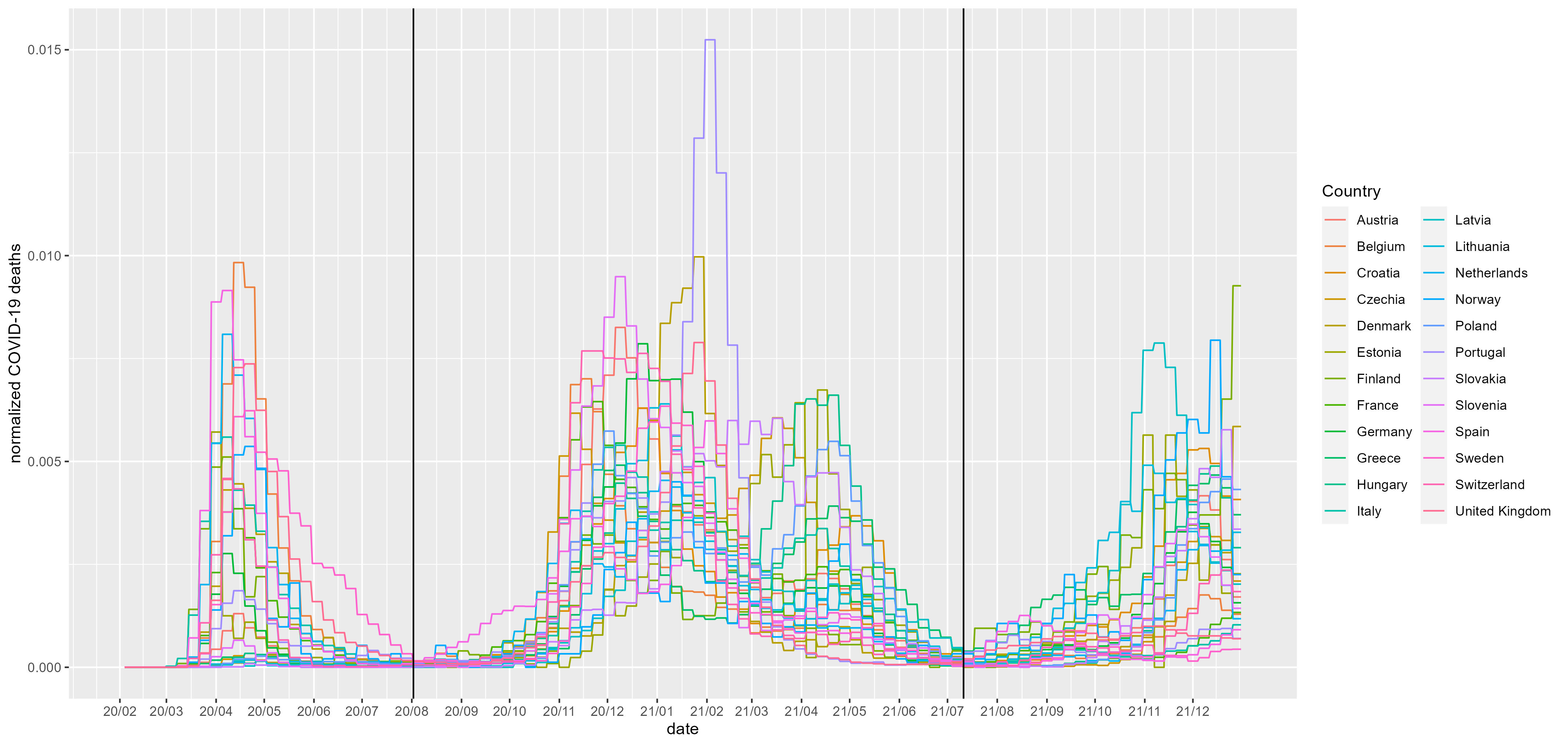}
\caption{Normalized \covid deaths per million for all 24 countries. Vertical lines indicate the boundaries of the pandemic phases.}
\label{gph:phases}
\end{figure}

In what follows, we explain in detail how pandemic waves were defined for the purpose of the reverse causality analysis of Subsection~\ref{sub:reverse}.
In order to define the beginning and the end of the pandemic waves in each country we start by dividing the period of analysis from 2020/02/04 to 2021/12/31 into three phases. The phases are the same for all countries. Here are the steps for defining the phases:
\begin{itemize}
\item[(i)] normalize the curves of \covid deaths per million by dividing the values by the total number of \covid deaths per million in the entire period of analysis;
\item[(ii)] add the normalized \covid deaths per million for all countries;
\item[(iii)] the boundaries of the phases are defined as the local minima of the curve obtained in step (ii).
\end{itemize}
Normalization ensures that low mortality countries are given non negligible weight in the sum. \autoref{gph:phases} shows the normalized \covid deaths per million for all countries and the division of the period of analysis into three phases. Phase~1 spans the period from 2020/02/04 to 2020/08/02, phase~2 from 2020/08/03 to 2021/07/11 and phase~3 from 2021/07/12 to 2021/12/31.

Once the phases are defined, for each country and each phase, a wave is defined as the smallest time interval contained in the phase that contains at least $99\%$ of the \covid deaths in that phase. This definition is inspired by the notion of highest posterior density intervals often used as credible intervals in Bayesian Statistics. Graphs showing pandemic waves and the period within waves corresponding to indices \texttt{maskbeginwave} and \texttt{maskpeakwave} are presented on the output files \texttt{Waves.pdf} and \texttt{WavesMarked.pdf} available on a GitHub repository \citep{Tausk2026}. All R code, datasets and output files used for the purpose of the reverse causality analysis of Subsection~\ref{sub:reverse} are also available on that repository. The analyses of Subsection~\ref{sub:twowayflawed} were done using obvious adaptations of the code made available by the authors of \citep{Cerqueira-Silva2026} on \citet{csthiago2025}.

\bibliographystyle{myauthordate4}
\bibliography{ref}

\end{document}